\documentclass[english,aps,prl,reprint]{revtex4-1}
\usepackage[latin9]{inputenc}
\usepackage{amstext}
\usepackage{amssymb}
\usepackage{graphicx}

\makeatletter

\@ifundefined{textcolor}{}
{%
 \definecolor{BLACK}{gray}{0}
 \definecolor{WHITE}{gray}{1}
 \definecolor{RED}{rgb}{1,0,0}
 \definecolor{GREEN}{rgb}{0,1,0}
 \definecolor{BLUE}{rgb}{0,0,1}
 \definecolor{CYAN}{cmyk}{1,0,0,0}
 \definecolor{MAGENTA}{cmyk}{0,1,0,0}
 \definecolor{YELLOW}{cmyk}{0,0,1,0}
}

\makeatother

\usepackage{babel}
\begin{document}

\title{Probing the Hotspot Interaction Length in NbN Nanowire Superconducting
Single Photon Detectors}

\author{J.J. Renema }

\altaffiliation{These authors contributed equally to this work}

\affiliation{Leiden Institute of Physics, Leiden University, Niels Bohrweg 2,
2333 CA Leiden, Netherlands}

\author{R. Gaudio}

\altaffiliation{These authors contributed equally to this work}

\affiliation{COBRA Research Institute, Eindhoven University of Technology, P.O.
Box 513, 5600 MB Eindhoven, Netherlands}

\author{Q. Wang}

\affiliation{Leiden Institute of Physics, Leiden University, Niels Bohrweg 2,
2333 CA Leiden, Netherlands}

\author{M.P. van Exter}

\affiliation{Leiden Institute of Physics, Leiden University, Niels Bohrweg 2,
2333 CA Leiden, Netherlands}

\author{A. Fiore}

\affiliation{COBRA Research Institute, Eindhoven University of Technology, P.O.
Box 513, 5600 MB Eindhoven, Netherlands}

\author{M.J.A. de Dood}

\affiliation{Leiden Institute of Physics, Leiden University, Niels Bohrweg 2,
2333 CA Leiden, Netherlands}
\begin{abstract}
We measure the maximal distance at which two absorbed photons can
jointly trigger a detection event in NbN nanowire superconducting
single photon detector (SSPD) microbridges by comparing the one-photon
and two-photon efficiency of bridges of different overall lengths,
from 0 to 400 nm. We find a length of $23\pm2$ nm. This value is
in good agreement with to size of the quasiparticle cloud at the time
of the detection event. 
\end{abstract}
\maketitle
Nanowire superconducting single photon detectors (SSPDs) \cite{Goltsman2001}
are a crucial technology for a variety of applications \cite{Natarajanrevie}.
These devices consist of a thin superconducting film which detects
photons when biased to a significant fraction of its critical current.
Although details of the microscopic mechanism are still in dispute\cite{Engelreview},
the present understanding of this process in NbN SSPDs is as follows
\cite{RenemaPRL,Vodolazov2012,Zotova2012,Engelpreprint,RenemaPRB,Luschewidthdep,ZotovaARXIV,VodolazovPRB,EngelarXiv2,RenemaNL,Kozorezov}:
after the absorption of a photon, a cloud of quasiparticles is created,
which is known as a hotspot. This cloud diffuses, spreading out over
some area of the wire. This causes the redistribution of bias current,
which triggers a vortex unbinding from the edge of the wire, if the
applied bias current is such that the current for vortex entry is
exceeded. The transition of a vortex across the wire creates a normal-state
region, which grows under the influence of Joule heating from the
bias current, leading to a measureable voltage pulse and a detection
event \cite{Kerman2006}.

Recently, applications of these detectors have been demonstrated or
proposed which rely the ability of such devices to operate as multiphoton
detectors, such as multiphoton subwavelength imaging \cite{Bitauld2010},
ultrasensitive higher order autocorrelation \cite{Zhou} and near-field
multiphoton sensing \cite{QiangOE}. These applications make use of
the fact that when biased at lower currents than required for single-photon
detection, the detector responds only when several photons are absorbed
simultaneously. This multiphoton response has moreover proven to be
of great significance in investigating the question of the working
mechanism of such devices.

For these multiphoton applications to work, the two photons must be
absorbed within some given distance of each other, which we will refer
to as the \emph{hotspot interaction length} $s$\emph{. }This length
determines the efficiency of an SSPD in the multiphoton regime: photons
which are absorbed far away from each other along the wire will not
be able to jointly cause a detection event, resulting in a reduction
of the two-photon detection probability. 

In this work, use this effect to measure the hotspot interaction length.
Our experiment is based on comparing the efficiency in the one-photon
and two-photon regime of a series of uniformly illuminated nanowires
of different lengths. We rely on quantum detector tomography \cite{Lundeen2008}
(QDT) to find the bias currents at which the one and two-photon regimes
occur. We experimentally find a hotspot interaction length of $s=23\pm2$
nm. We find that the tapers leading into our nanowires are photodetecting
over a length of approximately $35\pm6$ nm on each side. 

\begin{figure}
\includegraphics[width=9cm]{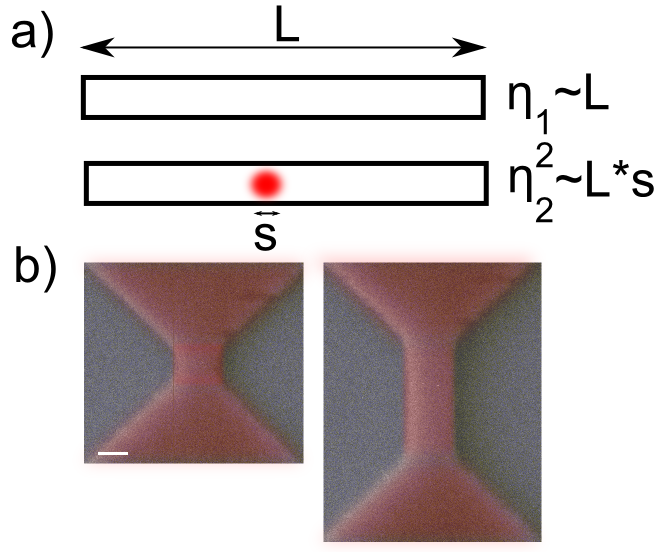}

\protect\caption{\emph{a) }Sketch of the experiment. Top pannel: a nanowire of length
$L$ is illuminated uniformly, and the current the nanowire is set
to be in the single-photon regime. Photon absorption at any point
in the wire is sufficient to cause a detection event. In the bottom
pannel, the detector is in the two-photon regime, and a detection
is observed only if the second photon is absorbed in the region (red
spot) where an excess quasiparticle concentration has been created
by the first photon. \emph{b) }False color SEM images of two nanowires,
of $L=100\ \mathrm{nm}$ and $L=400\ \mathrm{nm}$, respectively.}
\end{figure}

We interpret these results in terms of the diffusion-based vortex
crossing model of the detection event. We show that the measured hotspot
interaction length corresponds to the computed size of the quasiparticle
cloud at the moment of a detection event, which demonstrates the agreement
between our experiment and our numerical model. Finally, we discuss
the implications of these results for multiphoton-based SSPD applications. 

The detectors used in this experiment were patterned from a single
film (5 nm NbN on GaAs) to ensure that the properties of the wires
are as similar as possible. The film is deposited at a temperature
of 400 $^{\circ}$C, which was found to give the optimal critical
current for NbN on GaAs, and a film critical temperature of 9.6 K.
The detectors were patterned using conventional e-beam lithography
and reactive-ion etching in an $\textrm{SF}_{6}$ / Ar plasma. We
fabricated 16 detectors of each length, with lengths of L = 0, 100,
200 and 400 nm. The width of all detectors was nominally identical,
at 150 nm. 

For this experiment, it is crucial that the entire area of the detector
is active. It is known that the critical current of SSPDs shows variations\cite{Kerman2007};
NbN nanowires are inhomogeneous on a length scale below $100\ \mathrm{\mathrm{nm}}$
due to some sort of defects or intrinsic inhomogeneities, which manifest
themselves as a reduced value of the critical current \cite{RosalindaAPL}.
To avoid comparing dissimilar detectors, we measured the critical
current of our devices and selected one for each length with critical
currents between $27.4$ and $27.9\ \mathrm{\mu A}.$ This value is
consistent with earlier samples \cite{Bitauld2010,Renema2012,RenemaPRL,RosalindaAPL},
including bridge samples (nanodetectors) which have a very low probability
of containing a defect. 

To characterize these detectors optically, we perform QDT \cite{Lundeen2008,Lundeen,Renema2012,Qiangnoise}.
QDT relies on illuminating the detector with a set of known probe
states. By measuring the count rate as a function of bias current
and combining this with knowledge of the photon number distribution
in each probe state, we can measure the probablity of a detection
event given $n$ incident photons. Our modified tomography protocol
\cite{Renema2012} makes use of model selection to derive from the
observed counting statistics both a linear efficiency $\eta$ , and
a series of nonlinear parameters $\{p_{n}\},$ which correspond to
the detection probability of $n$ photons. 

We used a Ti:Saphhire laser with a wavelength of $\lambda=800\ \mathrm{nm}$
to perform detector tomography. This laser is well suited for this
experiment, because it has a pulse duration of approximately 100 fs.
In this way we avoid introducing the temporal response of the device
into the problem: our pulse duration is sufficiently short to act
as a delta-like excitation compared to all relevant timescales compared
to the lifetime of an excitation in an SSPD of a few tens of ps \cite{Zhou}.
The laser is attenuated by a $\lambda/2$ plate between two polarizers.
The second polarizer was set so as to maximize the count rate in the
device, which aligns the polarization of the light with the direction
of current flow, resulting in almost uniform illumination across the
wire \cite{RenemaNL}. The spot size was chosen to be much larger
than the length of the wire, to ensure uniform illumination along
the wire length.

\begin{figure}
\includegraphics[width=9cm]{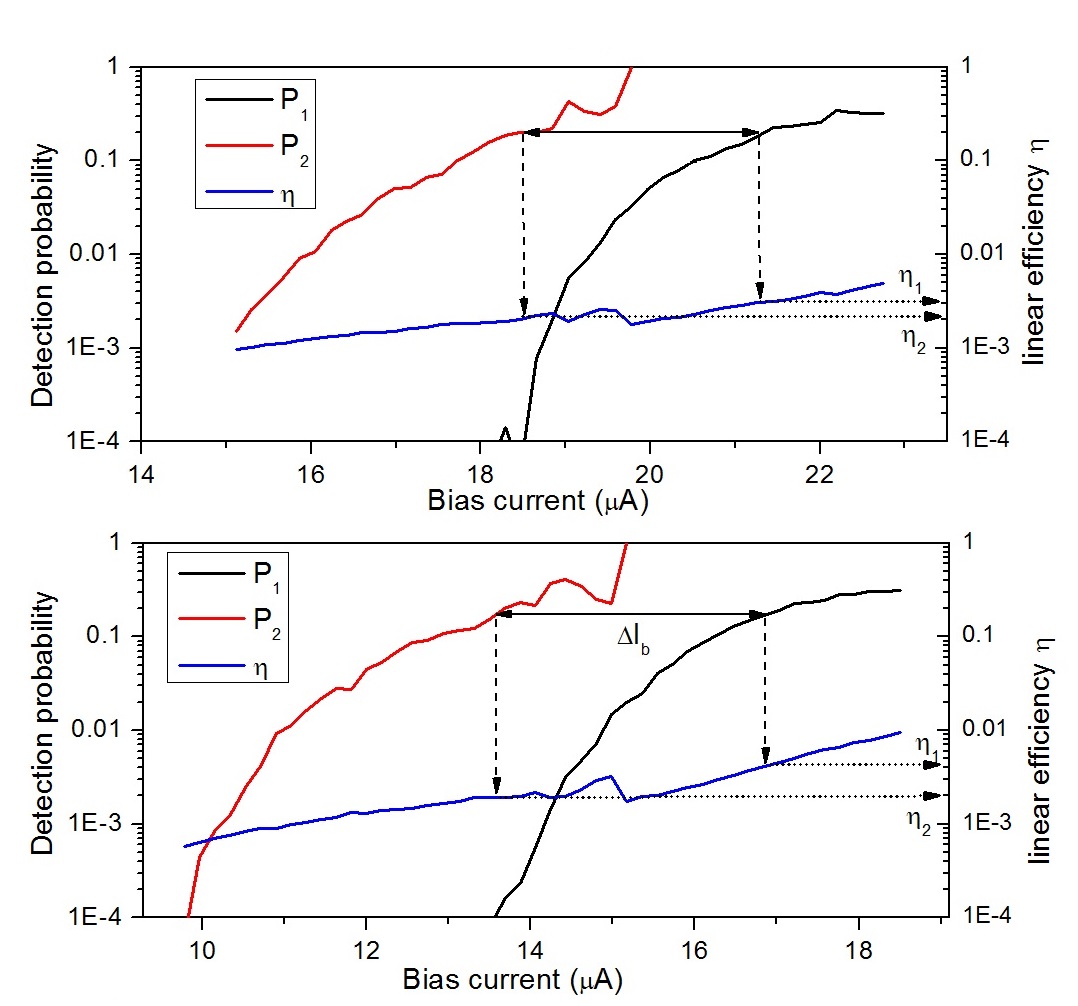}

\protect\caption{Full tomographic characterization of the $L=0$ nm sample \emph{(top)}
and the $L=400$ nm sample \emph{(bottom)}. The black and red lines
show the nonlinear detection probabilities for single photons ($p_{1})$
and photon pairs ($p_{2}$), respectively. The blue line shows the
linear efficiency $\eta.$ The dashed lines show how we obtain the
ratio of efficiencies in the one- and two-photon regime $\eta_{1}/\eta_{2}$. }
\end{figure}

Figure 2 shows two typical experimental results, for the L = 0 and
L = 400 nm wires. The results on these two devices are almost identical,
apart from the the linear efficiency parameter $\eta$, which falls
off faster for the longer wire. We conclude that the reduced detection
probability in the two-photon regime manifests itself as a reduction
in the linear efficiency. This is to be expected, considering that
one can interpret the reduction of the two-photon detection efficiency
geometrically: one of the two photons sets the area into which the
other has to be absorbed (see Figure 1). We therefore conclude that
the conjectured reduction in efficiency occurs and manifests itself
(in our parametrization) as a reduction of $\eta.$ 

To extract the hotspot interaction length from these measurements,
we consider the detection efficiencies in both photon regimes in detail.
For the one-photon regime, we expect $\eta_{1}=CL$, where $C$ denotes
the absorption probability per unit length of wire. For the two-photon
regime, we expect $\eta_{2}^{2}=C^{2}Ls,$ representing the fact that
the second photon has to be absorbed withing a distance $s$ from
the first%
\footnote{Throughout this work, all efficiencies are defined as single-photon
efficiencies.%
}. Therefore, we find: 
\begin{equation}
L/s=(\eta_{1}/\eta_{2})^{2}.
\end{equation}
We find that the change in $\eta$ is gradual with decreasing bias
current. Therefore, we compare equivalent points in the two photon
regimes. We start by finding the value of $\Delta I_{b}$ such that
$p_{2}(I_{b})=p_{1}(I_{b}+\Delta I_{b})$, as shown in Fig 2. We then
take the ratio of efficiencies $\eta_{1}/\eta_{2}=\eta(I_{b}+\Delta I_{b})/\eta(I_{b})$.
We find that for currents where $p_{1,2}\gtrsim0.15,$ the resulting
ratio is independent of bias current. This enables us to associate
one value of $\eta_{1}/\eta_{2}$ with each device. 

\begin{figure}
\includegraphics[width=9cm]{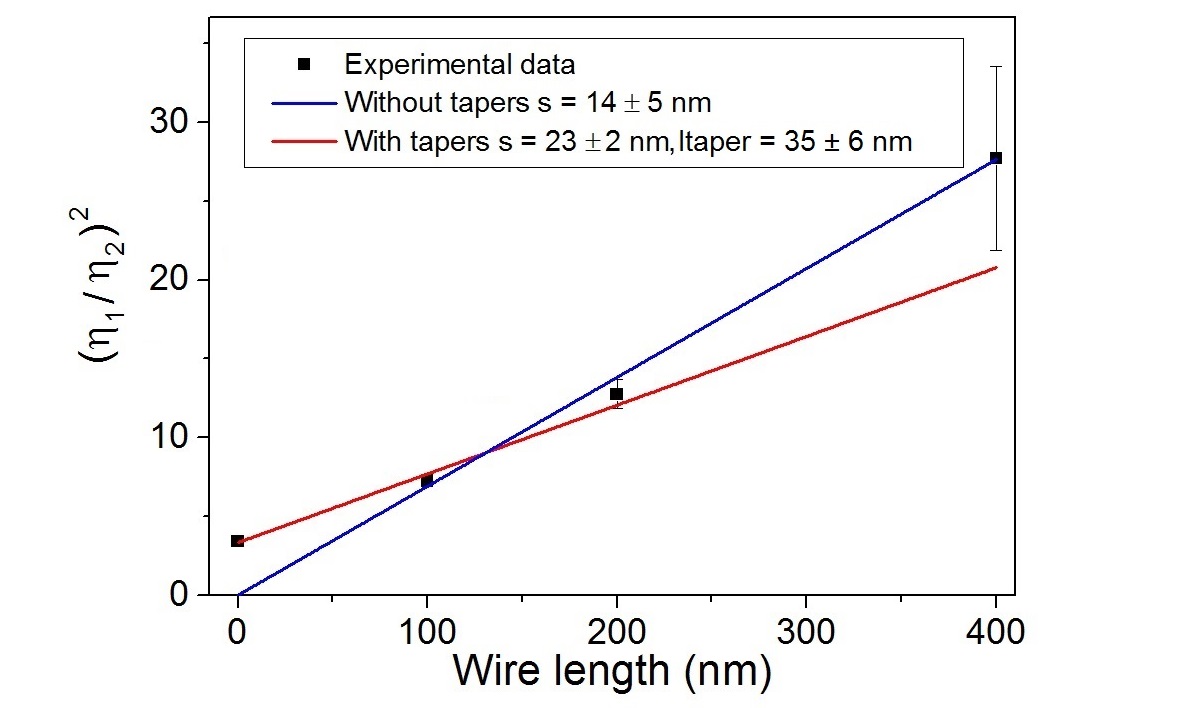}

\protect\caption{Ratio of linear efficiencies $\eta_{1}/\eta_{2}$ for the one and
two-photon regimes, derived from tomography as show in the previous
figure. The blue line shows a fit which does not take into account
photodetection events in the tapers leading to the wire. The straight
lines are fits to the data that either neglect (blue line) or include
(red line) an additional taper. From that fit, we find $s=23\ \mathrm{nm},$
$L_{taper}=36\ \mathrm{nm}.$ }
\end{figure}

Figure 3 shows the resulting values of $\eta_{1}/\eta_{2}$ for $L=0-400$
nm, from which we extract $s$. The point at $L=0$ is of note: for
this nanobridge, we find a value of $\eta_{\mathrm{1}}/\eta_{\mathrm{2}}=1.6$.
This is consistent with our earlier measurements\cite{Renema2012}
but not with our model, which would unphysically predict $\eta_{2}=\eta_{1}=0$,
as it neglects the possibility that photons absorbed close to the
end of the wire can trigger a detection. We model this effect by substituting
$L_{\mathrm{eff}}=L+2L_{taper},$ into eq. 1. Using this modified
wire length, we find values of $s_{hs}=23\pm2\ \mathrm{nm}$ and $L_{taper}=35\pm6\ \mathrm{nm}$.
The observed value of $L_{taper}$ is in reasonable agreement with
earlier estimates of the active area of such nanobridges, which found
$L_{taper}\approx50\mathrm{\ nm}$\cite{Bitauld2010,Renema2012}.

As a sanity check, we must consider that this experiment relies on
the assumption that the properties of the superconducting wire are
identical at both currents. To evaluate whether this is the case,
we consider the superconducting energy gap and the available number
of superconducting electrons. Both of these depend on the applied
bias current \cite{Tinkham,Anthore}. We find that the density of
superconducting electrons varies by 4\% over the range of currents
in which we performed this experiment, and that the superconcuting
gap varies by 1.4\%. Given that these values are much smaller than
the margin of error of our experiment, this justifies the assumption
of constant superconducting properties.

It should be noted that the efficiency changes smoothly across the
one and two-photon regime. In the simplest interpretation, one would
expect the linear efficiency to jump from $\eta_{1}$ to $\eta_{2}$
when the current is decreased from the one-photon regime to the two-photon
regime. However, as we have shown previously \cite{RenemaNL}, different
points along the cross-sections of the wire become photodetecting
at different bias currents. We conjecture that this effect leads to
the smearing-out of the transition between the one- and two-photon
regimes.

We note that the length scale which we have found is much smaller
than the width of the wire. We have 
assumed that photons which are absorbed at the same cross-section
of the wire are equivalent to a $d=0$ detection event, i.e. to a
photon with double the energy. We have observed previously\cite{RenemaPRL}
that a single photon of energy $2E$ has the same detection probability
as two photons of energy $E$. Our assumption is justified from theory: across
the wire, current continuity enforces an almost instantaneous\cite{Engelpreprint},
long-range interaction between the two hotspots. 

At this point, we have not attached any interpretation in terms of
detector physics to our observed length scale. To answer this question,
we perform a series of numerical simulations in COMSOL of current
continuity and quasiparticle diffusion, similar to those reported
on refs \cite{RenemaNL,Engelpreprint}. We have made three simplifications
compared to these references. First, we have approximated the process
of hot electron to QP conversion as an exponentially decaying source
of QP located at the photon absorption site. Second, we have ignored
the nonlinear interaction between the condensate velocity and the
number of quasiparticles, which amounts to taking the limit of low
quasiparticle densities, and equivalently, low photon energies. This
later approximation is somewhat justified by the fact that we are
in the regime where the energy-current relation is linear \cite{RenemaPRL}.
Finally, we only consider photon absorption events in the center of
the wire. 

\begin{figure}
\includegraphics[width=9cm]{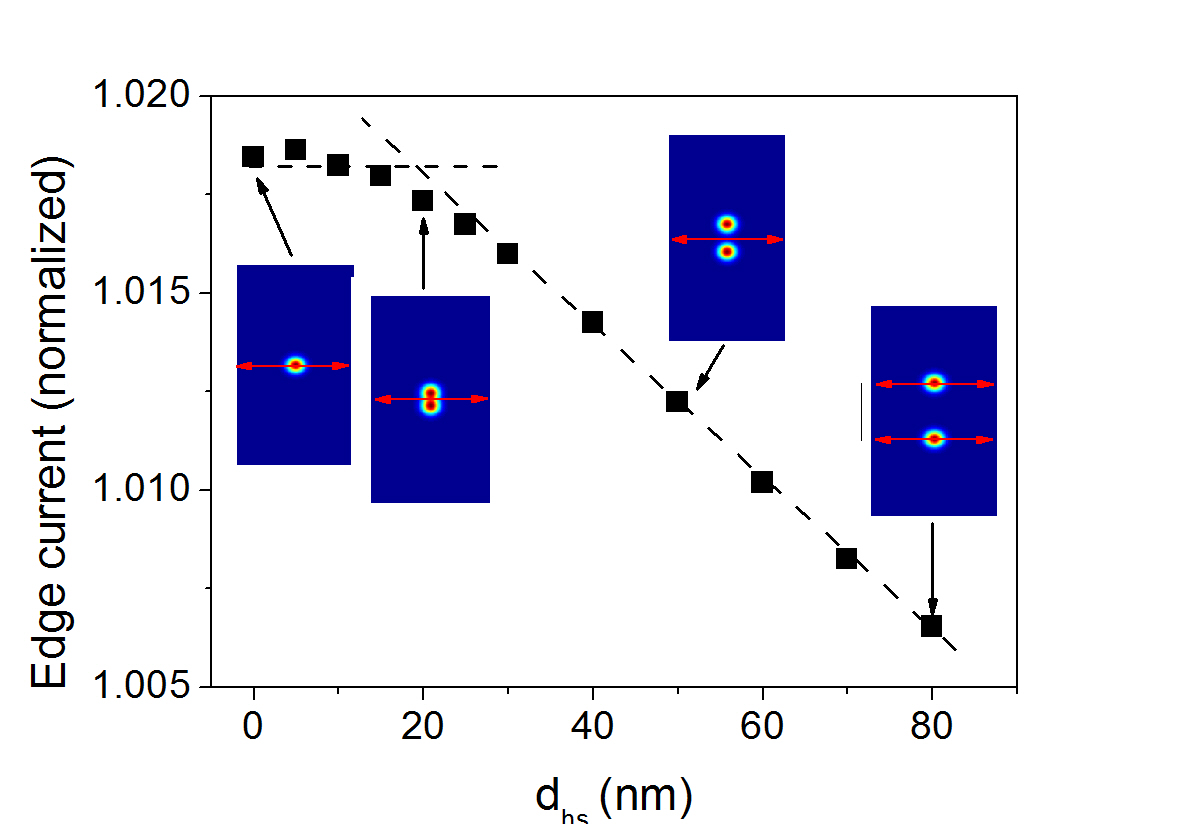}\protect\caption{Simulated edge current as a function of photon absorption separation,
normalized to the applied bias current. The dashed lines are guides
to the eye. The insets show the quasiparticle distribution at the
moment of maximum edge current, which we associate with the detection
event. The arrows indicate the point where the edge current is maximal. }
\end{figure}

Figure 4 shows that the length scale which we have measured in our
experiment is that of quasiparticle diffusion at the time of the detection
event. We plot the maximum value of the current along the edge of
the wire (in units of the applied bias current), which is the quantity
that is known to determine whether a detection event occurs \cite{Engelpreprint},
as a function of photon separation $d.$ By finding where and when
the current density along the wire is maximal, we can identify the
position and time of the photodetection event. In the four insets,
we plot the distribution of quasiparticles at the timestep when the
maximum edge current is achieved and we indicate the position where
this happens with a double red arrow. The edge current is roughly
constant up to $d_{hs}\approx20$ nm, and then starts to roll off.
The point where this rolloff happens occurs when the quasiparicle
clouds no longer significantly overlap. We therefore identify the
observed hotspot interaction length with the size of the QP cloud.
This result enables us to convert the observed length scale into a
timescale, since the diffusion constant for quasiparticles is known\cite{Engelpreprint}
to be $D=0.4-0.6\mathrm{\ cm^{2}/s}$. Using the relation $s=\sqrt{Dt}$,
we find a value of $t_{det}=2.7\mathrm{\pm0.6\ ps.}$ This is in good
agreement with the value predicted in ref \cite{Engelpreprint}. At
a range of $d$ = 20 - 60 nm, the two absorbed photons still interact
through the current continuity condition. Essentially, the current
crowding caused by the first QP cloud has not healed before the current
encounters the second QP cloud. However, this length scale is not
visible in Fig 4: the edge curent decreases smoothly from $d\approx20$
nm onwards. We therefore conclude that it is the length scale set
by the QP cloud and not by current crowding that determines the hotspot
interaction length. 

We discuss the implications of this result for the use of SSPDs in
multiphoton sensing. Increasing the length of the wire will increase
the probability of multiphoton events, but each $\sim$20 nm long
segment of the wire will essentially act as a seperate multiphoton
detector. This means that the efficiency of such devices will be low:
for a typical $100\ \mathrm{\mu m}$ long SSPD, the overall detection
probability in the two-photon regime would be $10^{-4}$ lower than
in the single-photon regime, and correspondingly for higher photon
regimes. This demonstrates that the only way to obtain highly efficient
multiphoton detection in SSPDs is to go to far-subwavelength focussing,
perhaps by the use of nano-antennas. 

Recently, a similar experiment was performed on WSi \cite{Stevenshotspots}.
In this experiment, it was found that the experimental data on two-photon
pump-probe measurements \cite{marsili2014hotspot,Marsillihotspots}
could be well explained by a static hotspot of $s>100\ \mathrm{nm}$.
It is possible that self-confinement of the hotspot plays a larger
role in WSi than in NbN due to the larger fraction of Cooper pairs
which are destroyed in the former material. This would be the first
evidence of a qualitative difference in the detection mechanism between
NbN and WSi SSPDs. 

Finally, we propose an application for the present work. If the experiment
presented here is performed on wires which are longer than the size
of the impinging optical beam, the diameter of the optical beam takes
the role of the wire length $L$. If the overall shape of the beam
is known (e.g. that it is Gaussian), this enables measurement of the
beam diameter with an accuracy far below the diffraction limit. 

In conclusion, we have observed that the size of an excitation in
NbN SSPDs is approximately 23 nm. We have shown that this number can
be interpreted as the size of the quasiparticle cloud at the moment
of detection. This observation is consistent with the predictions
of the diffusion-based vortex crossing model.
\begin{acknowledgments}
We thank Eduard Driessen, Alexander Kozorezov, Denis Vodolazov, Martin
Stevens, Francesco Marsili and Andreas Engel for useful discussions.
This work is part of the research programme of the Foundation for
Fundamental Research on Matter (FOM), which is financially supported
by the Netherlands Organisation for Scientific Research (NWO) and
is also supported by NanoNextNL, a micro- and nanotechnology program
of the Dutch Ministry of Economic Affairs, Agriculture and Innovation
(EL\&I) and 130 partners. JJR acknoledges support of Dirk Bouwmeester's
NWO Spinoza award. 
\end{acknowledgments}

\end{document}